\documentclass{elsart}
\newcommand{\be}{\begin{eqnarray}}
\newcommand{\ee}{\end{eqnarray}}

\usepackage[dvips]{graphicx}
\graphicspath{{./}}

\begin{document}

\begin{center}
{\LARGE Excitation of the GDR and the Compressional Isoscalar Dipole
State by $\alpha$ scattering } 
\end{center}

\bigskip
\begin{center}
J.A. Christley$^{\rm a)}$, E.G. Lanza$^{\rm b,e)}$, S.M. Lenzi$^{\rm c)}$, 
M.A. Nagarajan$^{\rm d)}$, and A. Vitturi$^{\rm c)}$ \\
{\em a) Department of Physics, University of Surrey, Guildford,
GU2 5XH, UK  \\
b) Dipartimento di Fisica and INFN, Catania, Italy\\
c) Dipartimento di Fisica and INFN, Padova, Italy \\
d) Department of Physics, UMIST, Manchester M60 1QD, UK\\
e) Departamento de F\'{i}sica Atomica, Molecular y Nuclear, Sevilla, Spain\\ }
\end{center}

\begin{center}
{\bf Abstract}
\end{center}

{\small The excitation of the isovector giant dipole resonance (GDR)
by alpha scattering is investigated as a method of probing the neutron
excess in exotic nuclei. DWBA calculations are presented for $^{28}$O
and $^{70}$Ca and the interplay of Coulomb and nuclear excitation is
discussed. Since the magnitude of the Coulomb excitation amplitude is
strongly influenced by the Q--value, the neutron excess plays an
important role, as it tends to lower the energy of the GDR.  The
excitation of the compressional isoscalar dipole state in $^{70}$Ca by
$\alpha$ scattering is also investigated. It is shown that the
population of this latter state may be an even more sensitive probe of
the neutron skin than the isovector GDR. }

\vspace{1cm}

\hspace{1cm} 

\bigskip

\noindent
\begin{tabbing}
Keyword list:  DWBA for $\alpha$ scattering on $^{28}$O, $^{70}$Ca, 
isoscalar and isovector 
\\
dipole resonance. 
\hspace{1cm} \= 
\end{tabbing}

\bigskip

\newpage

\bigskip


During recent years, there has been considerable interest in the study
of the structure of nuclei with large neutron excess~\cite{r1}.
Relativistic and non--relativistic self--consistent Hartree-Fock
Bogoliubov approaches~\cite{r2} to the structure predict that the
neutron and proton densities in the neutron-rich nuclei have different
shapes, with the neutron densities extending quite a bit further than
the proton densities. It had been pointed out by Clement, Lane and
Rook~\cite{r3} as well as by Satchler~\cite{r4} that the excitation of
the isovector giant dipole resonance by isoscalar hadronic interaction
becomes possible if the neutron and proton densities have different
shapes. There have been experiments with $\alpha$ particle probes to
test this idea~\cite{r5} and theoretical interpretation of the
excitation mechanism with a macroscopic model for the GDR has been
given by various authors~\cite{r6}.

\bigskip
\noindent
Isoscalar compressional dipole states (IDR) which are generated by the
operator $\sum_{i} z_i r_i^2$ were first considered by Harakeh and
Dieperink~\cite{r7} and by Van Giai and Sagawa~\cite{r8}. Recently,
the effect of the neutron excess on the GDR transition density and the
IDR transition density were studied within a Hartree-Fock plus RPA
with Skyrme interaction~\cite{r9,r9bis}. In the case of the GDR, it
was observed that the effect of neutron excess was to fragment the
strength distribution together with a shift of the centroid to lower
energy (cf. also~\cite{r10}). In the case of the IDR, a strong
concentration of the strength at low energy was predicted in nuclei
with neutron excess.

\bigskip
\noindent
In this note, we present the results of DWBA calculations of
$\alpha$-inelastic scattering related to the excitation of the GDR and
the IDR. We consider the very neutron rich nuclei $^{28}$O and
$^{70}$Ca in order to assess the effect of the neutron excess on the
cross sections. We also study the effect of the Q--value on the
Coulomb and nuclear amplitudes.
\begin{figure} 
\begin{center}
\includegraphics[bb= 110 110 550 719,angle=0,scale=0.6]{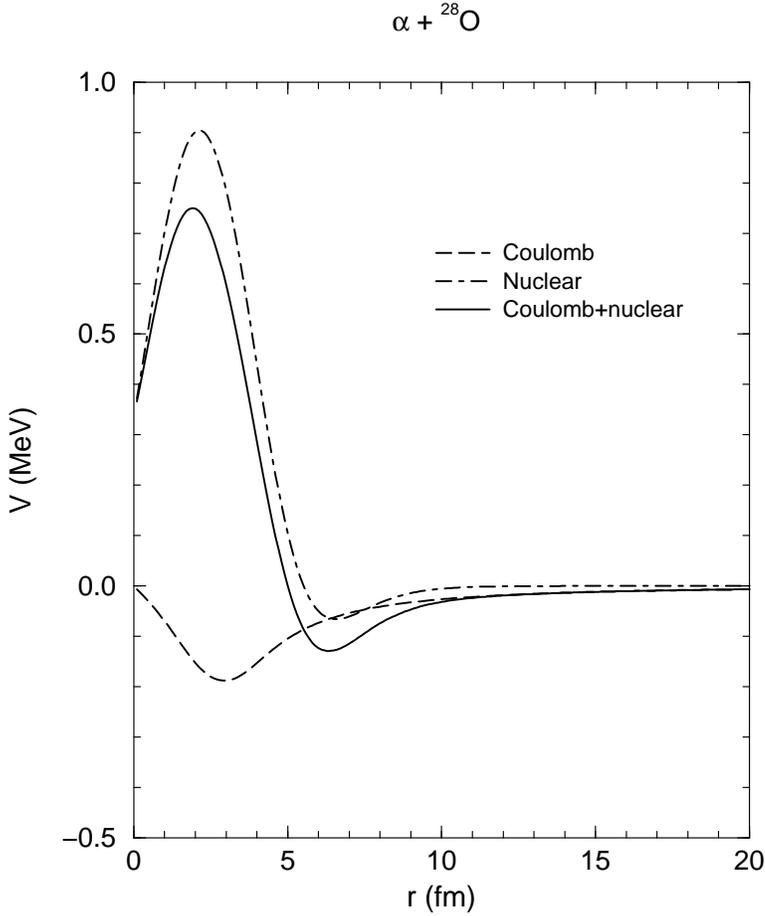}
\end{center}
\caption { Nuclear and Coulomb transition potentials for the
  excitation of the GDR in the $\alpha$+$^{28}$O system.}
\label{f1}
\end{figure}

\bigskip
\noindent
We first consider the excitation of the GDR in $^{28}$O by scattering
of alpha particles. The GDR is assumed to have an energy of 80
$A^{-\frac{1}{3}} \approx 26$ MeV.  The distorting potential for the
DWBA calculation was taken to consist of real and imaginary
Woods-Saxon form factors of depths $V_0$=29 MeV, $W_0=V_0/2$ with
the same radius and diffuseness parameters, $R_0$= 5.05 fm and $a_0$ =
0.63 fm~\cite{BW}. The hadronic and Coulomb transition potentials were
obtained by folding the RPA transition densities of Catara et
al.~\cite{r9} with the $\alpha$-nucleon and the Coulomb potentials,
respectively. They are shown in figure 1. Figures 2 and 3 show the
cross section and partial wave cross section for excitation of the GDR
in $^{28}$O.
\begin{figure} 
\begin{center}
\includegraphics[bb= 110 110 550 719,angle=-90,scale=0.55]{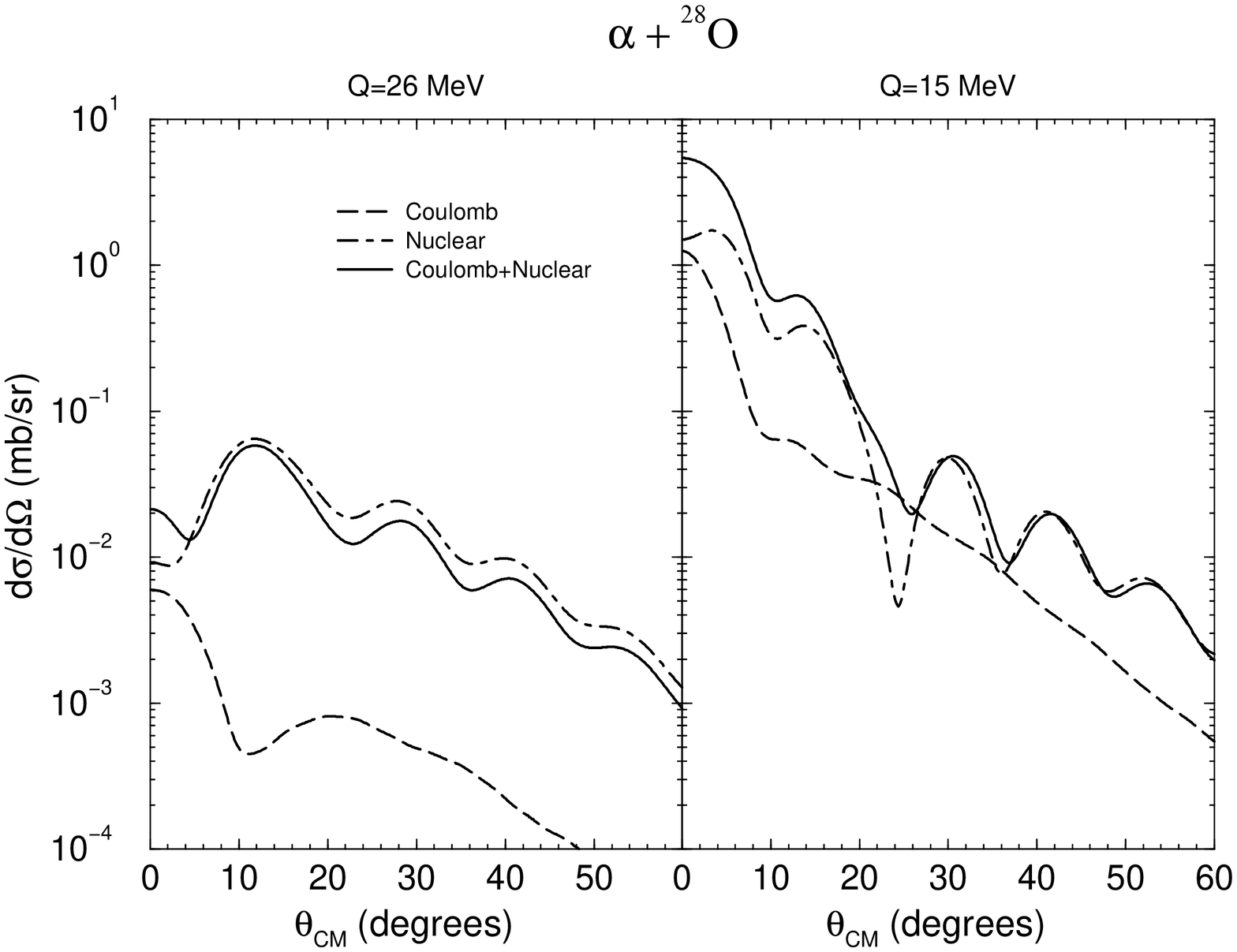}
\end{center}
\caption { GDR excitation cross section for $\alpha$ scattering
  at E$_{\rm L}$= 68 MeV on $^{28}$O for two different energy of the
  GDR: Q=26 MeV (left part) and Q=15 MeV (right part).}
\label{f2}
\end{figure}

\begin{figure} 
\begin{center}
\includegraphics[bb= 110 110 600 719,angle=-90,scale=0.55]{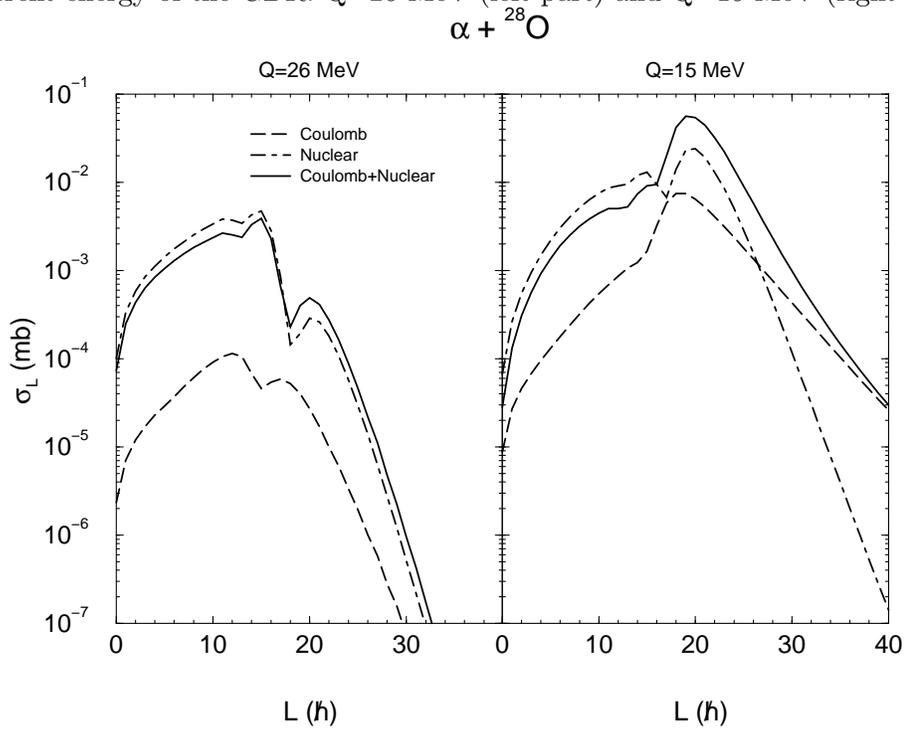}
\end{center}
\caption { Partial--wave cross sections for the same reaction of
Fig. 2: Q=26 MeV (left part) and Q=15 MeV (right part).}
\label{f3}
\end{figure}

\bigskip
\noindent
It is seen that in this case the low value of Z combined with the very
large excitation energy of the GDR strongly suppress the Coulomb
excitation cross section. The partial wave cross sections (left part
of figure 3) show that the nuclear contribution is dominant for all
partial waves. Still the smaller Coulomb contribution gives
appreciable interference effects.  As predicted in refs.~\cite{r4,r9},
due to the fact that the isoscalar dipole transition density $(\delta
\rho_n + \delta \rho_p)$ has different sign at small and large radii,
while the isovector one $(\delta \rho_n - \delta \rho_p)$ has a
definite sign, one finds a destructive Coulomb-nuclear interference
for small partial waves and a constructive one for large partial
waves.  As a result, one cannot observe the constructive
Coulomb-nuclear interference except at very small angles.

\bigskip
\noindent
One should note, however, that for neutron-rich nuclei, the average
excitation energy of the GDR is shifted to lower energies than that
given by the hydrodynamical value. In view of the strong Q-value
dependence of the Coulomb dipole excitation amplitude, it is necessary
to use the correct energy of the GDR. In the case of $^{28}$O, the
average excitation energy of GDR predicted by the RPA~\cite{r9} is
around 15 MeV. If we use this Q-value, the hadronic and Coulomb
amplitudes are more comparable in magnitude and become dominant at low
and high partial waves respectively.  One can then observe a strong
Coulomb-nuclear interference (right part of figure 2). This is also
apparent in Fig. 3 (right part) where we present the partial cross
section for the GDR at 15 MeV.

\bigskip
\noindent
We consider now the case of the excitation of dipole states in
$^{70}$Ca. Following the results of the RPA calculation~\cite{r9}, the
GDR for $^{70}$Ca is assumed to be at 13.7 MeV and the IDR at 9.4 MeV.
The corresponding hadronic transition potentials obtained by folding
these transition densities are shown in Fig.~4, together with the
contributions coming from the nuclear and Coulomb parts.  We note
that, although the total form factors have almost the same radial
shape the composition of the two is quite different: in the case of
the IDR the Coulomb contribution is smaller than in the GDR, while the
nuclear part is higher.  As a consequence, in the IDR case the form
factor is given, in the peripheral region around 10 fm, essentially by
the nuclear contribution, which is clearly an effect to be related to
the neutron excess.  The distorting potential for the DWBA calculation
were obtained by double folding the M3Y potential~\cite{sat} with a
gaussian $\alpha$--particle density and the Hartree-Fock
$^{70}$Ca~\cite{r9} density. The imaginary part of the potential has
been assumed to be half the real part.  The excitation cross section
for the GDR is shown in the right part of figure 5, while the left
part of the same figure shows the corresponding cross section for
exciting the IDR. In both cases the same bombarding energy of the
$\alpha$--particle was chosen (E$_{\rm L}$= 40 MeV).  As can be seen
in the figures, whereas both the Coulomb and the hadronic excitations
contribute to the GDR, the contribution from nuclear field is strongly
enhanced in the case of the excitation of the IDR. The larger cross
section obtained for the excitation of the IDR with respect to the GDR
can be strictly related to their form factors (Fig. 4). Effects due to
the smaller excitation energy may also play a role. Since the low
energy peak of the IDR is originated by the presence of the neutron
skin, its excitation results in a direct probe of neutron excess.
\begin{figure} 
\begin{center}
\includegraphics[bb= 110 110 550 719,angle=-90,scale=0.55]{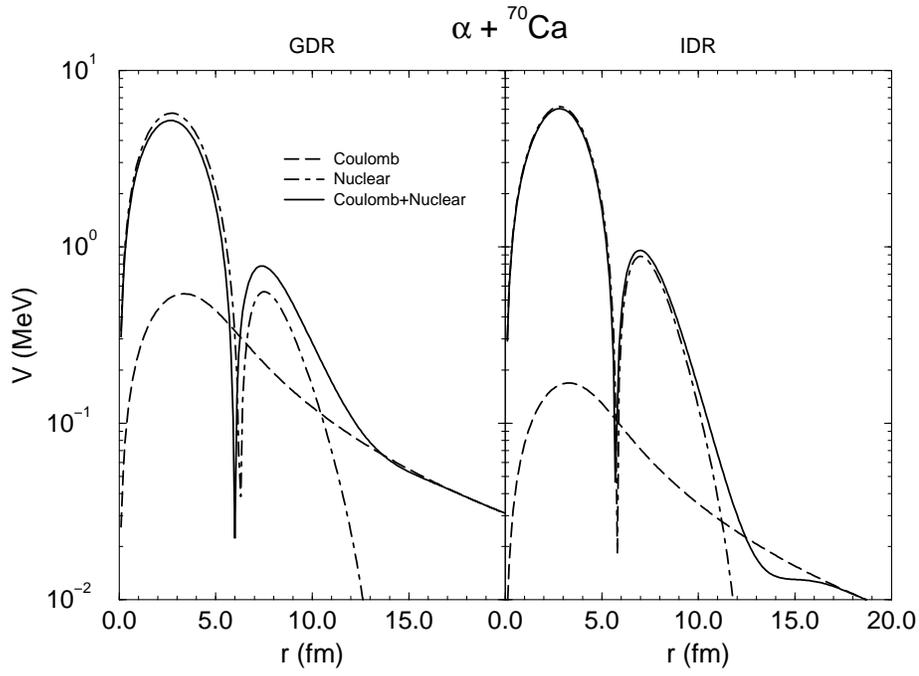}
\end{center}
\caption { Transition potentials for the excitation
  of the GDR (right) and IDR (left) in the $\alpha$+$^{70}$Ca system.}
\label{f4}
\end{figure}

\begin{figure} 
\begin{center}
\includegraphics[bb= 110 110 620 719,angle=-90,scale=0.55]{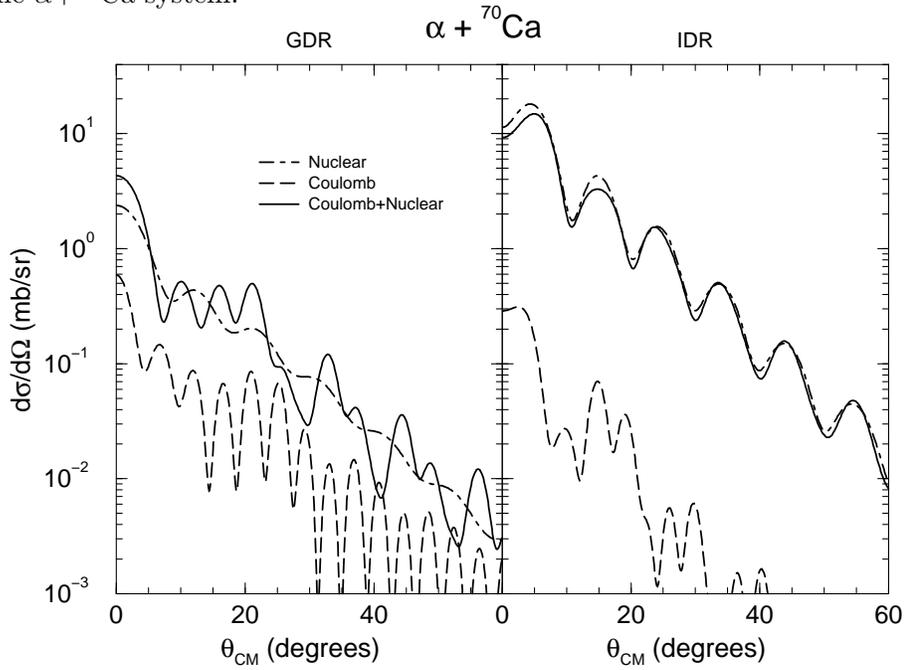}
\end{center}
\caption { Cross section for the excitation of the GDR (right)
  and IDR (left) in $^{70}$Ca in the reaction $\alpha$+$^{70}$Ca at
  E$_{\rm L}$ = 40 MeV. The single contributions from nuclear and
  Coulomb excitations are shown together with the total result.}
\label{f5}
\end{figure}

\bigskip
\noindent
To summarize, we note that the nuclear excitation of GDR by hadronic
probes becomes more feasible in neutron-rich nuclei. In these nuclei,
the GDR strength is strongly fragmented with some of the strength
extending to lower excitation energy.  Both Coulomb and nuclear
excitations are strongly affected by the Q-values favoring lower
Q-values. The variation of the amplitudes with the Q-value is so
strong that even though the low energy components of the GDR carry a
small fraction of the B(E1) strength, this is more than compensated by
the enhanced amplitudes. A second type of dipole state which can be
strongly excited by isoscalar probes is the compressional isoscalar
dipole state. In neutron-rich nuclei, the corresponding strength
distribution is fragmented, pushing a large fraction of the strength
to lower excitation energies. Thus, these states are amenable to
strong excitation by isoscalar probes, providing a direct sensitivity
to neutron excess.

\vfill
\end{document}